# Exploring Correlation Methods
# to Determine QCD $\beta$ Functions on the Lattice

Gunnar S. Bali, Christoph Schlichter[*]

*Physics Department, Bergische Universität, Gesamthochschule Wuppertal*

*Gaußstrasse 20, 42097 Wuppertal, Germany*

Klaus Schilling

*HLRZ c/o KFA, 52425 Jülich Germany*

*and DESY, Hamburg, Germany*

(August 25, 1995)

## Abstract

We investigate — as an alternative to usual Monte Carlo Renormalization Group methods — the feasibility of extracting nonperturbative QCD $\beta$ functions directly from a lattice analysis of correlations between the action and Wilson loops. We test this correlation technique numerically in four dimensional $SU(2)$ gauge theory, on a $16^4$ lattice at $\beta = 2.5$ and find very promising results.

# I. INTRODUCTION

Determinations of QCD $\beta$ functions beyond perturbation theory are becoming more and more important. This holds in particular for the lattice laboratory, where the quality of data allows for increasingly detailed physics investigations, like scaling analyses and thermodynamic studies in anisotropic lattices.

Much work has been done in the past to extract $\beta$ functions from the response of lattice observables to changes in the parameters (coupling and, in the case of full QCD, quark masses) [1–4]. In these analyses, the $\beta$ function is typically retrieved from finite difference measurements, at fixed values of physical observables (like physical hadron masses). In practice one has to operate at sufficiently different values of the parameters, however, i.e. not really locally.

It is well known in statistical mechanics that response functions are related to correlations and therefore can be determined without recourse to such finite difference methods. In lattice gauge theories this correspondence has become known in the context of lattice sum rules [5–7]. The latter have been considered in the past as mere consistency checks for the underlying (action and energy) distribution functions, assuming the $\beta$ function to be known. Given the recent progress in lattice technology, however, we think it is timely to turn things around and exploit sum rules for a nonperturbative determination of the $\beta$-function.

In this paper, we will test this approach. For simplicity we will work with SU(2) gauge theory. We shall extract the $\beta$ function, $B_\beta = \partial \beta / \partial \ln a$, from the action-density in presence of a static quark-antiquark pair, ie. from correlators between (smeared) Wilson loops and plaquettes. The method can readily be generalized to other response functions, such as $B_\kappa = \partial \kappa / \partial \ln a$, appearing in full QCD.

In the next section we will introduce the idea of the method for the case of dynamical fermions. This includes of course the pure gauge theory case that we pursue in the subsequent numerical part. The details of the numerical techniques are explained in section 3. The results are compared to results from interpolation methods in the concluding section 4.

## II. THE GIST OF THE METHOD

We consider the (isotropic) lattice action

$$S = S_g(\beta; U) + S_f(\kappa; U, \psi, \bar\psi) \quad, \tag{1}$$

where $S_g$ ($S_f$) denotes the gluonic (fermionic) part. The most common form of $S_g$ is the Wilson action

$$S_g = \beta S_W = \beta \sum_{n,\mu>\nu} S_{\mu\nu}(n) \tag{2}$$

with

$$S_{\mu\nu}(n) = 1 - \frac{1}{N} \text{Re Tr}\left(U_\mu(n) U_\nu(n+\hat\mu) U_\mu^\dagger(n+\hat\nu) U_\nu^\dagger(n)\right) \tag{3}$$

where $N$ denotes the number of colors.

The fermionic action is given by:

$$S_f = \sum_{m,n} \bar\psi_m M_{mn} \psi_n \tag{4}$$

with the fermionic matrix

$$M = \mathbf{1} - \kappa D. \tag{5}$$

$\bar\psi$ and $\psi$ are to be understood as vectors in color, Dirac, and position spaces, $M$ as a $4NV \times 4NV$ matrix. The hopping term, $D$, corresponds to the $\bar\psi \partial_\mu \gamma^\mu \psi$ part of the continuum Dirac action. The results presented here are quite general and do not depend on the specific form of the action.

Let us now illustrate the basic idea for computing the $\beta$ functions. We consider the expectation value of a (gluonic) operator, $O$,

$$\langle O \rangle = \frac{\int dU d\psi d\bar\psi \, O e^{-S(U,\psi,\bar\psi)}}{\int dU d\psi d\bar\psi \, e^{-S(U,\psi,\bar\psi)}}, \tag{6}$$

and compute the responses under changes in the couplings $\beta$ and $\kappa$:

$$C_\beta[O] = \frac{\partial \ln\langle O\rangle}{\partial \ln \beta} = -\left(\frac{\langle OS_g\rangle}{\langle O\rangle} - \langle S_g\rangle\right) \tag{7}$$

and

$$C_\kappa[O] = \frac{\partial \ln\langle O\rangle}{\partial \ln \kappa} = \left(\frac{\langle O\bar\psi\psi\rangle}{\langle O\rangle} - \langle\bar\psi\psi\rangle\right) \ . \tag{8}$$

Note that the chiral condensate, $\langle\bar\psi\psi\rangle = \langle\mathrm{Tr}M^{-1}\rangle$, can be obtained from inversion of the fermionic matrix on fixed gluonic background fields.

The expectation values of $S_g$ and $\bar\psi\psi$ are to be understood without normalization, i.e. without division by factors $6V$ or $4NV$. Eq. (8) has been obtained from the identity

$$\left\langle\frac{\partial S}{\partial \ln \kappa}\right\rangle = -\kappa\langle\bar\psi D\psi\rangle = \langle\bar\psi M\psi\rangle - \langle\bar\psi\psi\rangle = 4NV - \langle\bar\psi\psi\rangle \ . \tag{9}$$

If the derivatives of $\ln O$ in respect to the couplings can be related to derivatives of dimensionful quantities, the right hand sides (RHS's) of Eqs. (7) and (8) can be measured to determine the latter. Let us consider as our test operator $O$ a Wilson loop $W(A)$, encircling a (rectangular) area $A$. In the confined phase one obtains the area law

$$W(A) \propto \exp(-KA) \ , \tag{10}$$

where $A$ has to be kept sufficiently large to suppress perimeter terms and (in case of dynamical fermions) sufficiently small for string breaking to be ineffective. The constant $K$ is related to the physical string tension, $\sigma = K(\beta,\kappa)a^{-2}(\beta,\kappa)$. This relation allows to determine the lattice resolution $a$.

From Eq. (10), one obtains:

$$\frac{1}{A}\frac{\partial \ln W(A)}{\partial \ln \gamma} \approx -\sigma\frac{\partial a^2}{\partial \ln \gamma} = -2K\frac{\partial \ln a}{\partial \ln \gamma} \ , \tag{11}$$

where $\gamma = \beta$ or $\kappa$. Finally we find from Eqs. (7) and (8),

$$B_\beta = \frac{\partial \beta}{\partial \ln a} = \beta\left(\frac{\partial \ln a}{\partial \ln \beta}\right)^{-1} \approx -2AK\beta C_\beta^{-1}[W(A)] \tag{12}$$

and

$$B_\kappa \approx -2AK\kappa C_\kappa^{-1}[W(A)]. \tag{13}$$

Note, that the relations Eqs. (11) and (12) have been derived by Karsch long ago [8], in the context of thermodynamics.

## III. NUMERICAL ANALYSIS

Though, in theory, Eqs. (12) and (13) offer the possibility for a direct measurement of the $\beta$ functions, in practice, strong fluctuations of the correlators, in particular for large Wilson loops, might affect the applicability of this method. We can demonstrate, however, that appropriate use of smearing and noise reduction techniques in conjunction with the exploitation of the known subasymptotic form of Wilson loops (from transfer matrix methods) leads to substantial improvement in correlation signals, and provides success.

We start from the spectral decomposition of a rectangular Wilson loop with a spatial extent of $R$ and a temporal extent of $T$ lattice units:

$$W(R,T) = C(R)e^{-V(R)T}\left(1 + \sum_{i \geq 1}\frac{C_i(R)}{C(R)}e^{-\Delta V_i(R)T}\right), \tag{14}$$

where $\Delta V_i(R) = V_i(R) - V(R)$ are gaps between ground state and excited state potentials. The overlaps $C(R)$ and $C_i(R)$ are positive, with their sum normalized to one ($C(R) + \sum_i C_i(R) = 1$). Manipulation of the spatial parts of the Wilson loop leaves the eigenvalues $V$ and $V_i$ unaffected but alters the amplitudes $C$ and $C_i$.

Let us differentiate a smeared Wilson loop (with fixed lattice $R$, $T$) with respect to $\beta$:

$$\beta^{-1}C_\beta[W(R,T)] = \frac{\partial \ln\langle W(R,T)\rangle}{\partial \beta} \tag{15}$$

$$= \frac{\partial \ln C(R)}{\partial \beta} - T\left(\frac{\partial V(R)}{\partial \beta} + \frac{C_1(R)}{C(R)}\frac{\partial \Delta V_1(R)}{\partial \beta}e^{-\Delta V_1(R)T}\right) + \cdots . \tag{16}$$

Ground state enhancement implies an additional suppression of the (exponential) correction terms to the linear behaviour, which is welcome. Therefore, we construe a combination of suitable spatial paths to deplete excited states contributions up to a few percent, by use of APE-like smearing methods [9,10].

As a side product to our previous work on flux distributions [10], we have carried out simulations on a $16^4$ lattice at $\beta = 2.5$, with 8680 configurations[1]. In Fig. 1, we illustrate the numerical situation for $R = 3$. We observe a clean linear behaviour of $C_\beta(T)$, starting out from $T$ as low as $T_{min} = 2$, which is indicative for the suppression of excited states. Note that the slope is – apart from the normalization $2K$ – our signal for $-\frac{\partial V(R)}{\partial \beta}$.

After having extracted $\partial V(R)/\partial\beta$ from linear fits in $T \geq T_{\min}$, starting from reasonable values of $T_{\min}$, we separate the potential into a self energy part from the static propagator — which will diverge in the continuum limit — and a physical part,

$$V(R) = V_0 + V_{ph}(R) = V_0 + av(aR). \qquad (17)$$

$v(r)$ is the potential, measured in physical units, and does not depend on the lattice spacing. Starting from this separation ansatz, one finds[2]

$$\frac{\partial V(R)}{\partial \beta} = \left[V_{ph}(R) + RV'_{ph}(R)\right] \frac{\partial \ln a}{\partial \beta} + \frac{\partial V_0}{\partial \beta} \quad . \qquad (18)$$

By applying a two parameter fit to the RHS in accord with the expected behavior, $B_\beta$ can be extracted.

In the confined phase of QCD, the potential can be fitted to the ansatz $V_{ph} = KR - e/R$ with high accuracy, starting from $Ra \approx 0.3$ fm. From this parametrization, one expects the behaviour

$$\frac{1}{2K} \frac{\partial V(R)}{\partial \beta} = c + R B_\beta^{-1} \qquad (19)$$

with a constant $c$ that amounts to $1/(2K)$ times the derivative of $V_0$ with respect to $\beta$. It is nice to see at this point, that the Coulomb like term has disappeared, due to its scale invariance[3]. Indeed, this will enable us to perform the analysis down to very small values of $R$.

---

[1] For the present feasibility study we have not attempted to average over all possible orientations and locations of Wilson loops nor have we made use of link integration techniques. The cost of the method will readily be reduced by a factor 30–50, once this is done.

[2] The $RV'$ term has been missing in the original form of the sum rules [5], as has been noticed in ref. [11].

The actual numerical situation is depicted in Fig. 2, where the values for $-\frac{1}{2K}\frac{\partial V(R)}{\partial \beta}$ are plotted versus $R$. The data indeed exhibits the linear behaviour in $R$ predicted in Eq. (19), with a slope

$$B^{-1}_{\beta=2.5} = -4.03 \pm 0.23 \quad , \tag{20}$$

for $R_{\min} = 3$. Table I collects the results from various fits, with different cuts. We find the value of the $\beta$ function to be fairly stable under reasonable changes of $R_{\min}$ and $T_{\min}$. Note that the nonperturbative result differs appreciably from the perturbative two loop expectation for the $\beta$ function which reads $B^{-1} = -2.53$ for $SU(2)$ gauge theory at $\beta = 2.5$.

## IV. DISCUSSION

It is interesting to compare our local value for $B^{-1}$ with previous nonperturbative results that are based on interpolations to $a(\beta)$. A fit to the string tension values across the region $2.3 \leq \beta \leq 2.85$ renders the estimate [10] $B^{-1} \approx -2.94$, whereas the analysis of the critical temperature data presented by Engels *et al.* [12] led to the the value $B^{-1} \approx -3.21$ for $\beta = 2.5$. The difference between these "global" estimates might be interpreted as a 10 % interpolation uncertainty.

On the other hand we find the result from our present local method, Eq.(20) to differ by more than two standard deviations from the "global" estimates. A high statistics scan through the $\beta$-region with the local method is obviously needed to study the situation in more detail and settle the issue.

It looks rather promising to pursue this method in the case of *full* QCD, as knowledge about the $\kappa$-response function will be very helpful in guiding full QCD simulations through coupling parameter space. The same holds for the finite temperature field theory with anisotropic couplings.

---

[3] We thank H. Dosch for discussions on this point.

We would like to stress that the local method will be particularly efficient in a high dimensional coupling space, compared to finite difference methods that combine different Monte Carlo runs.

For the purpose of our feasibility study we have used the string tension, being the most precise dimensionful quantity accessible in lattice simulations [13]. It would be interesting to benchmark the power of correlation methods on other observables.

## V. ACKNOWLEDGEMENTS

During completion of this work we enjoyed discussions with H. Dosch, C. Michael, and H. Rothe. We thank the DFG for supporting the Wuppertal CM-2 and CM-5 projects (grants Schi 257/1-4 and Schi 257/3-2) and the HLRZ for computing time on the CM-5 at GMD. We appreciate support by EU contracts SC1*-CT91-0642 and CHRX-CT92-0051.

FIGURES

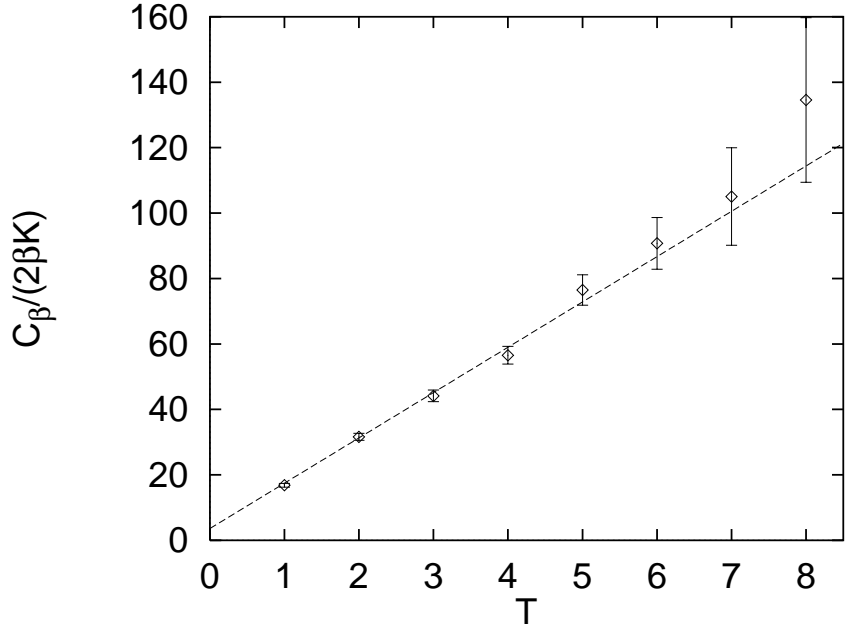

FIG. 1. The correlator $C_\beta[W(R,T)]/(2\beta K)$ versus $T$ at $\beta = 2.5$ and $R = 3$. The slope at large $T$ is proportional to the derivative of the potential in respect to $\beta$.

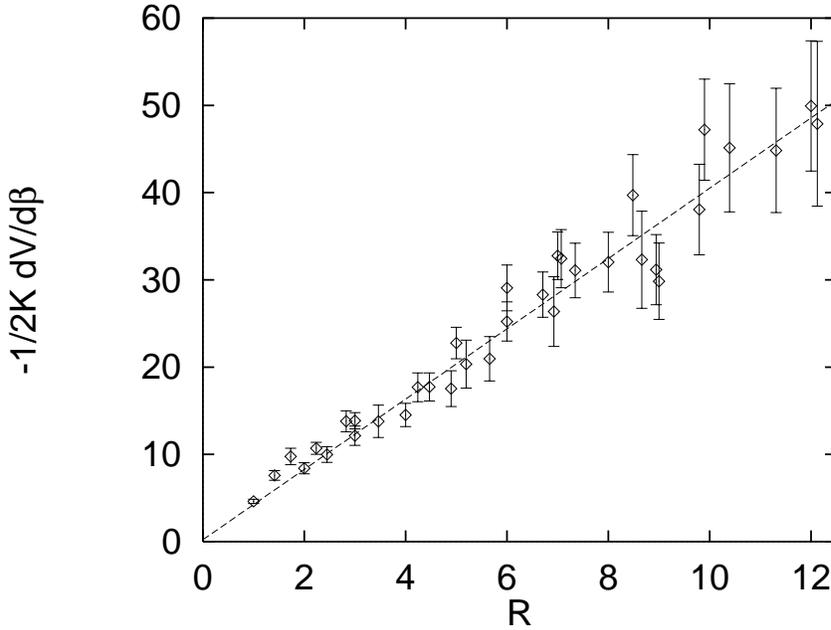

FIG. 2. The combination $-\frac{1}{2K}\frac{\partial V(R)}{\partial \beta}$ versus $R$ at $\beta = 2.5$. The slope at large $R$ is the inverse of the $\beta$ function, $B_\beta^{-1}$.

TABLES

TABLE I. Results on the $\beta$ function $B$ and the constant $c$ (Eq. (19)) at $\beta = 2.5$ for different cuts $T_{\min}$ and $R_{\min}$.

| $T_{\min}$ | $R_{\min}$ | $-B_\beta^{-1}$ | $-c$ |
| --- | --- | --- | --- |
| 2 | 2.45 | 4.0(2) | 0.7(8) |
| 2 | 2.83 | 3.9(2) | 1(1) |
| 2 | 3.00 | 4.0(2) | 0(1) |
| 2 | 3.00 | 4.2(2) | -1(1) |
| 2 | 3.46 | 4.2(3) | -1(2) |
| 2 | 4.00 | 4.3(3) | 1(2) |
| 2 | 4.24 | 4.1(3) | 1(2) |
| 2 | 4.47 | 4.1(4) | 1(2) |
| 2 | 4.90 | 4.0(4) | 1(3) |
| 2 | 5.00 | 3.7(4) | 1(3) |
| 3 | 2.45 | 4.1(4) | 0(2) |
| 3 | 2.83 | 3.9(4) | 2(2) |
| 3 | 3.00 | 4.1(4) | 1(2) |
| 3 | 3.00 | 4.4(5) | -1(2) |
| 3 | 3.46 | 4.2(6) | 0(3) |
| 3 | 4.00 | 4.3(6) | -1(2) |
| 3 | 4.24 | 3.8(6) | 3(4) |
| 3 | 4.47 | 3.9(7) | 2(4) |
| 3 | 4.90 | 3.7(7) | 4(5) |
| 3 | 5.00 | 3.0(8) | 9(6) |

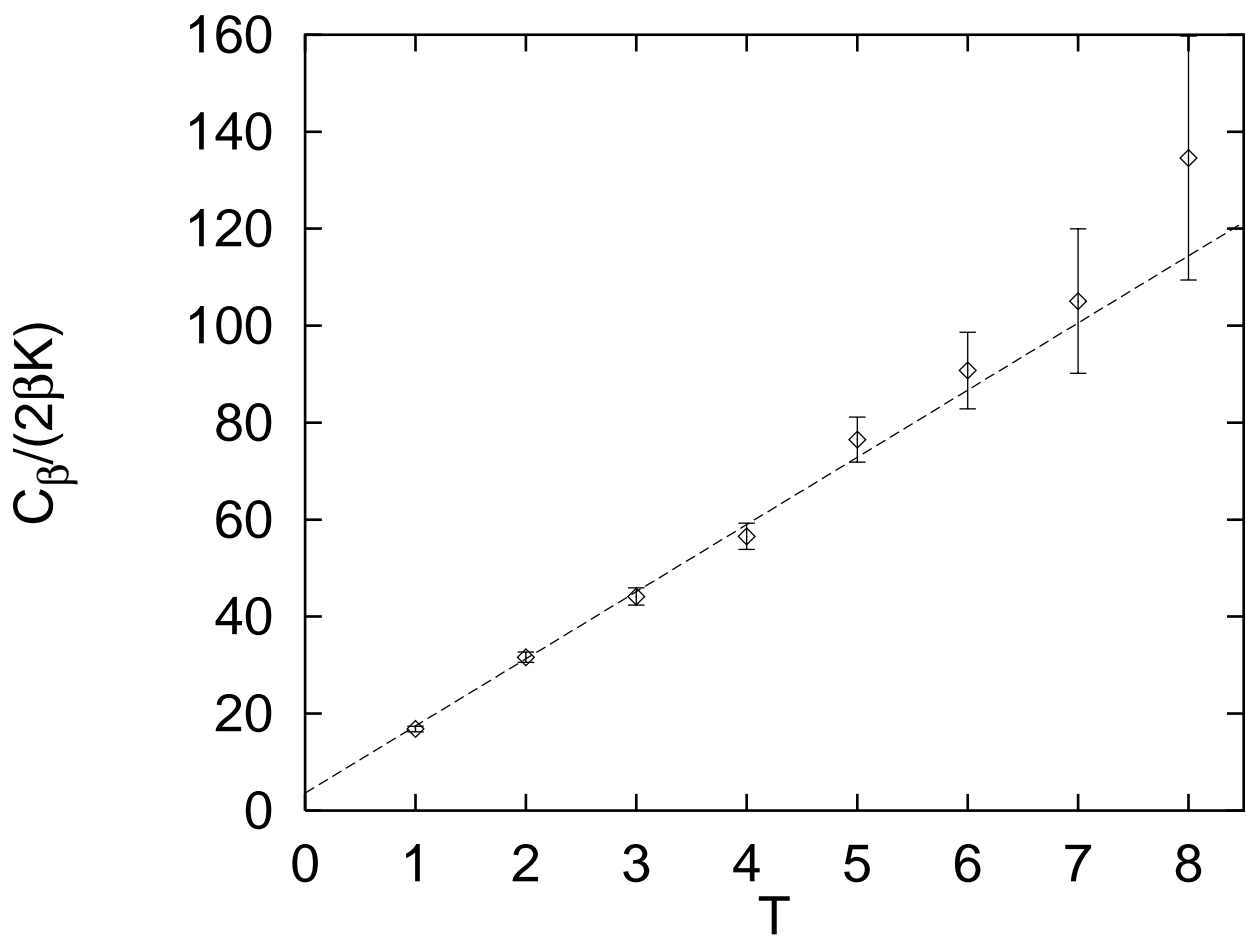

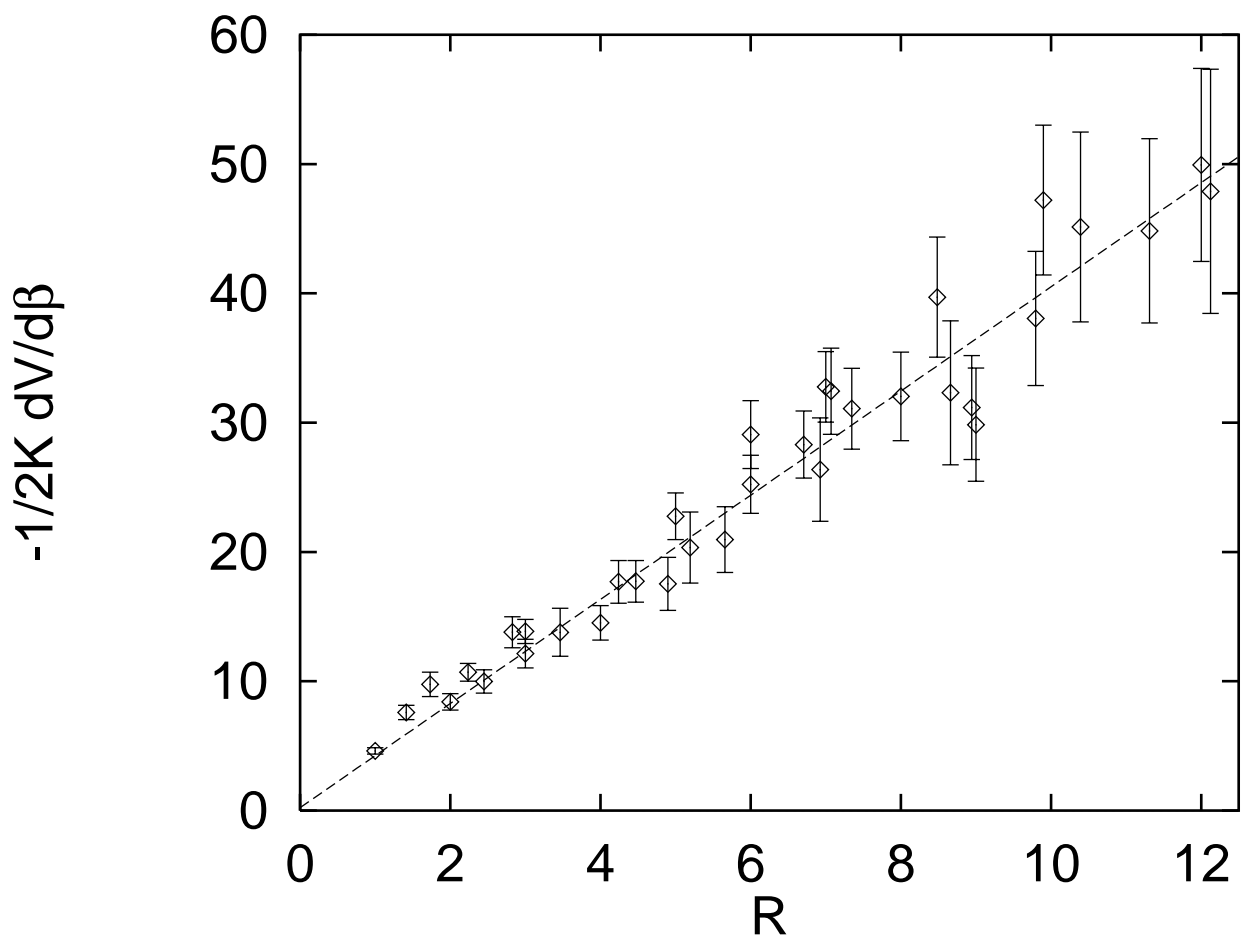